\documentclass[11pt]{article}
 \usepackage{jcappub} 
\usepackage{amsmath} 
\usepackage{graphicx,verbatim}
\usepackage{epsfig,multicol,bbm}

\newcommand\fverb{\setbox\fverbbox=\hbox\bgroup\verb}
\newcommand\fverbdo{\egroup\medskip\noindent%
			\fbox{\unhbox\fverbbox}\ }
\newcommand\fverbit{\egroup\item[\fbox{\unhbox\fverbbox}]}
\newbox\fverbbox

\newcommand{\InterpMC}{{\sc InterpMC}}
\newcommand{\CosmoMC}{{\sc  CosmoMC}}

\title{Caching and Interpolated Likelihoods:  Accelerating Cosmological Monte Carlo Markov Chains
}

\author[a,b]{Adam Bouland}
\author[a]{Richard Easther}
\author[a,c]{Katherine Rosenfeld}
\affiliation[a]{Department of Physics, Yale University, New Haven CT 06520, United States of America.}
\affiliation[b]{St. John's College, Cambridge, CB2 1TP, United Kingdom.}
\affiliation[c]{Harvard-Smithsonian Center for Astrophysics MS-10, 60 Garden Street, Cambridge, MA 02138, United States of America.}

\emailAdd{adam.bouland@aya.yale.edu}
\emailAdd{richard.easther@yale.edu}
\emailAdd{krosenfeld@cfa.harvard.edu}

\abstract{ We describe a novel approach to accelerating Monte Carlo Markov Chains. Our focus is cosmological parameter estimation, but the algorithm is applicable to any problem for which the likelihood surface is a smooth function of the free parameters and computationally expensive to evaluate. We generate a high-order interpolating polynomial for the log-likelihood using the first points gathered by the Markov chains as a training set. This polynomial then accurately computes the majority of the likelihoods needed in the latter parts of the chains.   We implement a simple version of this algorithm as a patch (\InterpMC) to \CosmoMC\ and show that it accelerates parameter estimatation by a factor of between two and four for well-converged chains. The current code is primarily intended as a ``proof of concept'', and we argue that there is considerable room for further performance gains.   Unlike other approaches to accelerating parameter fits, we make no use of precomputed training sets or special choices of variables, and \InterpMC\  is almost entirely transparent to the user.  }
 

\begin{document}
\maketitle
\section{Introduction}

\indent
Cosmological parameter values are typically estimated using Monte Carlo Markov Chains [MCMC].  MCMC techniques are far more efficient than a brute force exploration of a parameter space with a realistic number of independent variables. Despite this, running Markov Chains for a broad range of parameter combinations --  standard procedure when analyzing the cosmological implications of a major astrophysical dataset -- remains computationally expensive. Consequently,  algorithmic improvements that significantly increase the efficiency of this scheme without reducing its functionality are well worth exploring. 

Schematically, MCMC parameter estimation begins with a model, or {\em prior\/} which has a set of free parameters  \cite{Christensen:2001gj,Knox:2001fz,RubinoMartin:2002rc,Lewis:2002ah,Verde:2003ey}.  A likelihood function (derived for a  specific combination  of datasets) returns the relative probability that the ``observed sky'' was produced by the prior with a specific set of parameter values. After picking an initial point in the parameter space, we compute the likelihood at a new set of parameter values. The chain will {\em update\/} to this new point with probability,
\begin{equation}
\alpha(\Theta_n, \Theta_{n+1}) =  \text{min} \left\{ 1, \frac{{\cal L}(\Theta_{n+1})q(\Theta_{n+1}, \Theta_n)}{{\cal L}(\Theta_n)q(\Theta_n, \Theta_{n+1})} \right\}
\end{equation}
where ${\cal L}(x)$ is the relative likelihood of parameter vector $x$ and $q(x, y)$ is the proposal density  from $x$ to $y$.   Within  cosmology, \CosmoMC\  is a canonical and widely used implementation of the algorithm \cite{Lewis:2002ah}.  

A single likelihood can be computed in seconds, but a full set of chains requires the evaluation of many individual likelihoods.  Computing ${\cal L}$ is a nontrivial task since we must generate the  CMB angular power spectrum (or $C_\ell$) corresponding to our chosen parameter vector. Further, improvements in both angular resolution and signal-to-noise in future datasets will require $C_\ell$  to be computed  with increased precision and over a larger range of $\ell$ than currently necessary, sharply increasing the computational cost.  Hence there is a strong need to accelerate the MCMC analysis of cosmological data.

A number of improvements to standard MCMC parameter estimation have been proposed  to speed up or bypass the likelihood calculation. These include  {\sc CMBFit} \cite{Sandvik:2003ii},  a polynomial fit to the WMAP 1-year likelihood,  DASH \cite{Kaplinghat:2002mh}, which uses a combination of precomputation and analytic approximation, and WARP \cite{Jimenez:2004ct}, which  combines interpolation with a careful choice of orthogonal variables \cite{Kaplinghat:2002mh} to accelerate the computation of the power spectrum. The most mature packages are {\sc CosmoNet} \cite{Auld:2007qz} which trains a neural network to provide likelihood values, and PICO \cite{Fendt:2006uh}, which uses a large, precomputed training set.  

MCMC estimation can be used with a vast range of problems.  However, cosmological likelihoods -- particularly those derived from the Cosmic Microwave Background [CMB] experiments -- have two useful properties: they tend to be smooth functions\footnote{Exceptions to this rule certainly exist \cite{Easther:2004vq}, but it will not be a particularly burdensome restriction.} of the input parameters $x$ and evaluating them is computationally expensive.  Consequently, we attempt to speed parameter estimates by  caching computed values of the likelihood and then using interpolation to replace subsequent calls to the likelihood function within the MCMC code.  The smoothness of the likelihood ensures that the interpolated likelihood will be a good approximation to the actual value.   The computational overhead required by the caching and interpolation is small, compared to the cost of evaluating the likelihood directly, thus increasing the overall efficiency of the MCMC chains. 

Crucially, we make no use of precomputation when constructing our interpolation.  Our technique essentially works because the MCMC analysis itself ``discards" enough information to reconstruct an interpolated likelihood; there is no need for a training set.  This approach is straightforward computationally  -- we use a stock interpolation routine and make relatively small changes to \CosmoMC\ itself.  We implemented this algorithm as a patch to \CosmoMC, dubbed  \InterpMC. We find that without any serious attempts to optimize the interpolation the runtime required for a typical parameter estimation is reduced by a factor of 2 to 4 for well-converged chains, with no degradation in the results.  This improvement is nontrivial, although not as dramatic as that achieved  by methods that rely on precomputation or analytic approximations. These can improve the runtime of MCMC code by an order of magnitude or more, but often at the cost of a good deal of extra work (whether analytic or computational), which must be performed beforehand.  Further,   \InterpMC\, works for any  combination of datasets and cosmological model, is almost entirely transparent to the user, and offers a good deal of scope for future improvement. The source is available as a patch file to \CosmoMC.\footnote{{\tt http://easther.physics.yale.edu/interpmc.html} - at this point the code is offered as ``proof of concept", rather than production-ready code, but it has proved to be robust in a wide variety of settings.}

The structure of this paper is as follows. In Section~\ref{sec:overview} we describe our algorithm, and the details of its implementation. Performance  metrics and tests used to validate the interpolation scheme are discussed in Section~\ref{sec:test}. We  test \InterpMC\ with a variety of datasets (WMAP, ground based CMB, supernovae, BAO) and scenarios, including both the usual concordance cosmology, and models with curvature, neutrinos, running and tensors, along with associated run-times.  Finally, in Section~\ref{sec:disc} we summarize our work, and identify enhancements to \InterpMC\ that could further improve its performance.

\section{\InterpMC: Description and Implementation \label{sec:overview}}

The key ingredient of \InterpMC\ is a polynomial fit to computed likelihood values: this is  a function of a given model's free parameters. The interpolation data is gathered ``on the fly'': the first 10-30$\%$ of the chains typically yield enough points to construct an accurate interpolation. We use likelihoods computed for both accepted and unaccepted points, and sets of chains run in parallel  pool their interpolation data. Note that when using multiple datasets (e.g. BAO + WMAP7) we interpolate the {\em full\/} likelihood, rather than just the WMAP likelihood.

We find that  it is effective to fit the log-likelihood to an $n$-th order polynomial in the free parameters, rather than the likelihood itself. The likelihood is roughly Gaussian near the peak, and non-Gaussian corrections to the likelihood tend to be multiplicative, rather than additive. Consequently, while the likelihood and log-likelihood can both be expanded as polynomials, fitting to the log likelihood yields more stable results.    Further, we  exclude points from the interpolation which differ from the maximum log-likelihood value by more than a user-defined threshold, ensuring that the interpolation is not dominated by points that are infrequently visited by the Markov chains.

\subsection{Constructing the Interpolation}

\begin{figure}[tpb]
\epsfig{file= 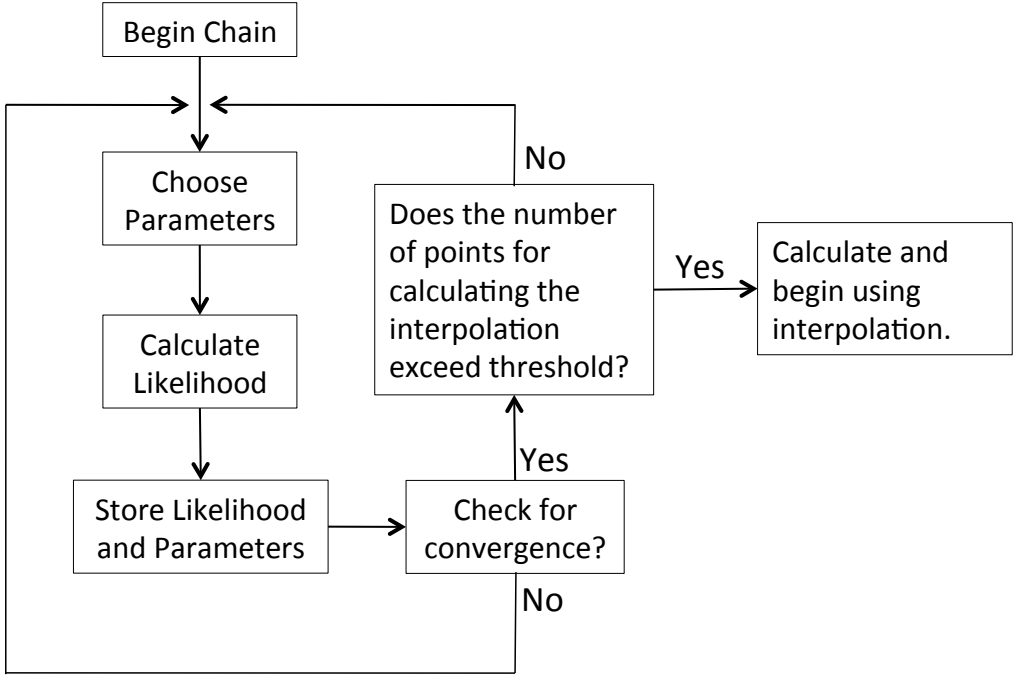,width=14cm} 
        \caption{Flow chart illustrating how the interpolation is constructed. }
	\label{fig:constructinterp}
	\end{figure}

The $n$-th order interpolating polynomial for the log-likelihood contains all combinations of the $P$ free parameters in the chains, up to $n$-th order.   Consequently, the interpolation algorithm is required to estimate the parameters $c_{ij\cdots k}$ in the following polynomial
\begin{equation}
\log{\cal L}(x_1,\cdots x_P) \approx c_0 + \sum_{i=1}^P c_i x_i + \sum_{i=1,j\ge i }^P c_{ij}  x_i x_j  +
\sum_{i=1,k\ge j, j\ge i}^P c_{ijk}  x_i x_j  x_k +  \cdots 
\end{equation}
where the expansion will be truncated at order $n$.   The variables $x_i$ are rescaled to have mean of zero and a standard deviation of unity, but are otherwise proportional to the ``raw'' variables in the chains. 

As pointed out below, we make no claim that our fitting procedure is optimal, but do demonstrate that it is good enough to  consistently improve the runtime of MCMC parameter estimates.    The polynomial has  $(P+n)!/{P!n!}$ unique coefficients, a number which obviously grows quickly  with both $n$ and $P$.  However,  we were able to fit a 4-th order polynomial with 9 free parameters while substantially accelerating the corresponding parameter estimation.  Within \CosmoMC, the convergence of a set of chains is tested periodically.  \InterpMC\ uses these same checkpoints to determine whether the number of collected points (across all chains) passes a threshold value set by the user, which we typically chose to be three times the number of free parameters in the polynomial.  When the threshold is exceeded, the chains share their databases of likelihoods. We first determine the peak likelihood encountered up to that point. The pooled points are then normalized and reduced to a subset whose log-likelihoods differ from the peak log-likelihood by less than some user-defined threshold.  We then use an unweighted least-squares algorithm to fit both $n-1^\textrm{th}$ and $n^\textrm{th}$ order polynomials to the cached likelihoods.   By doing so, we can estimate the accuracy of the fit: if the two polynomials differ substantially, the interpolation is dominated by the highest order terms, and cannot be relied upon.\footnote{We use a version of Applied Statistics algorithm AS 174 for the fitting \cite{1992Miller}.}

In addition to the free parameters in the chains, \CosmoMC\ returns a number of derived parameters. Some of these, such as the age of the universe,  only require the background FRW solution and can be trivially computed. Conversely, quantities such as  the clustering parameter $\sigma_8$ or the alternative tensor:scalar ratio $r_{10}$  require information obtained by running CAMB, which is precisely the step \InterpMC\ is designed to avoid.  Consequently, we also construct interpolations for these ``slow'' derived quantities. Given the quality of current data, these interpolations are extremely accurate since $\sigma_8$ and $r_{10}$ vary more slowly  than the likelihood, as functions of the free parameters.    A flowchart outlining the  use of the interpolation is shown in Figure \ref{fig:constructinterp}. 
	
\subsection{Using the Interpolation}

\begin{figure}[tb]\epsfig{file=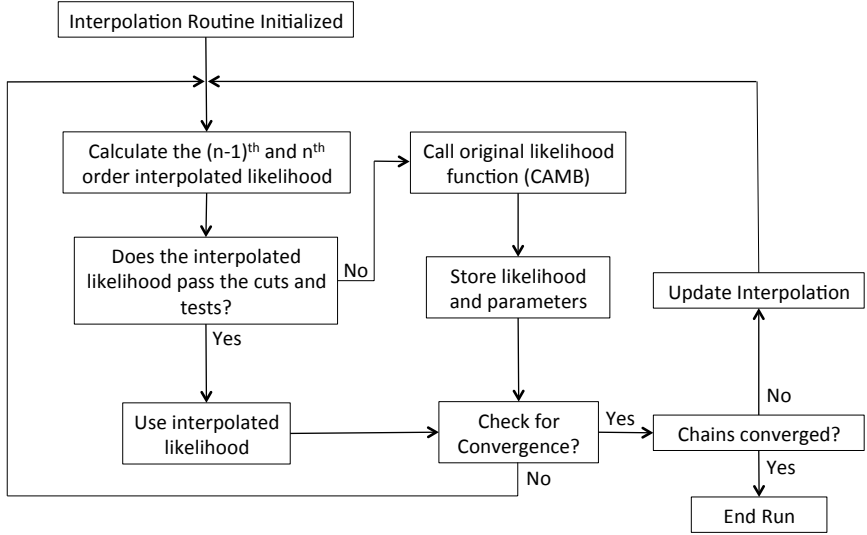,width=14cm}
		\caption{Flow chart illustrating how the interpolation is applied during the \CosmoMC\ run.  }
	\label{fig:useinterp}
	\end{figure}

\begin{figure}[tb]\epsfig{file=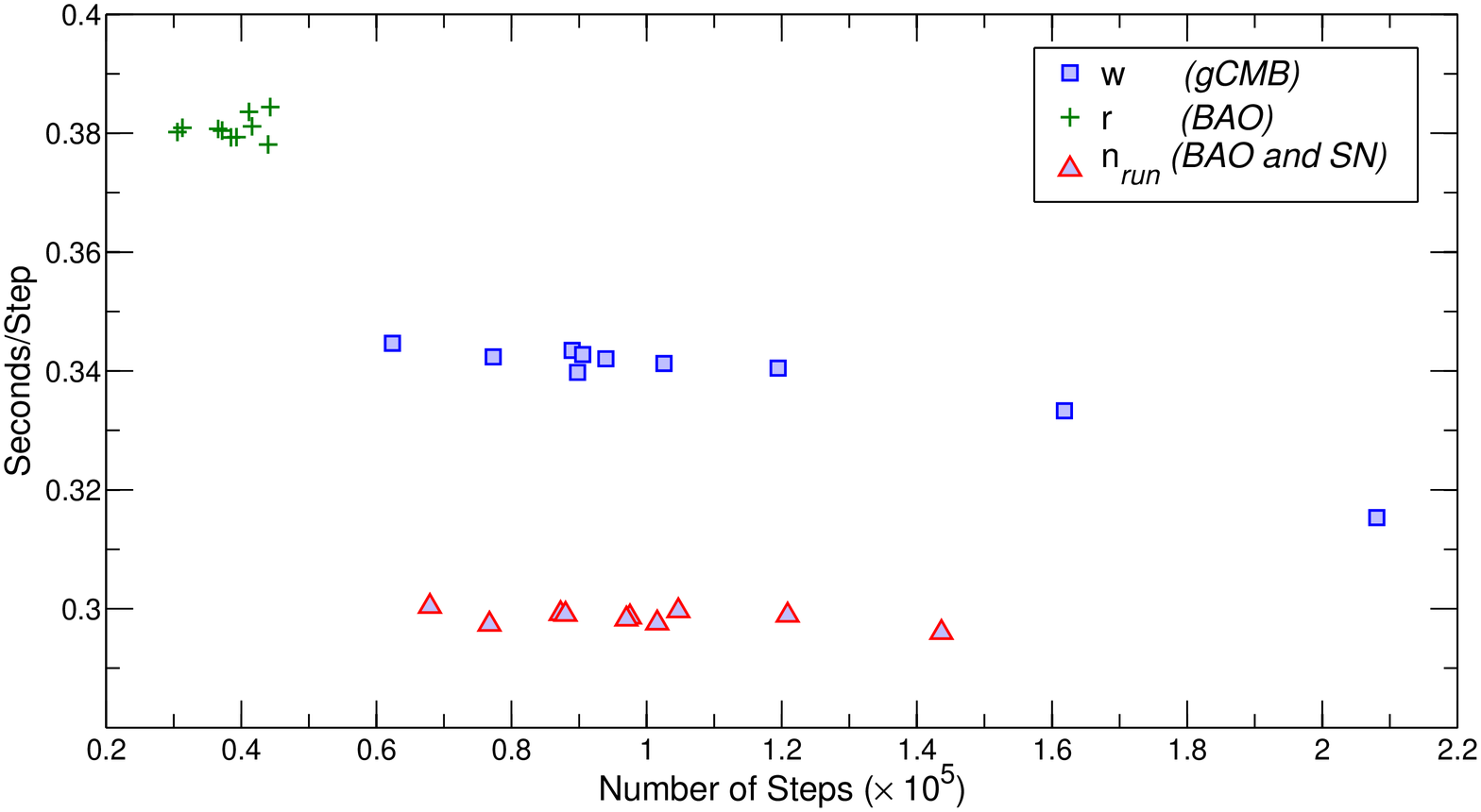,width=14cm} 
        \caption{We plot the  ratio of wallclock time to the number of accepted steps for ten different chainsets, for three different estimations.  For a given model, the total number of steps varies by a factor of a few for a given model but the ratio of runtime to the number of steps is roughly constant. The models shown here are a) WMAP + ground based CMB for $\Lambda$CDM $+ w$,  b) WMAP+BAO with $\Lambda$CDM $+ r$ and c)  WMAP + BAO + Supernovae with  $\Lambda$CDM $+ n_{\rm run}$.  }
	\label{fig:allthree}
	\end{figure} 

Once the polynomial coefficients have been calculated, the chains continue to run. As noted above, several criteria must be met for the interpolated likelihood to be used.  First, the interpolated likelihood must lie above the user-defined threshold, setting the maximum difference between the interpolated value and the peak value. Secondly, the $n-1^\textrm{th}$ and $n^\textrm{th}$ order values are compared, and must agree to within a specified tolerance.   When all these conditions are met, the chain obtains the log-likelihood from the $n^{\textrm{th}}$ order polynomial, and then accepts or rejects the corresponding step as usual.    If these conditions are not satisfied, the full likelihood code is called.  If the regular likelihood code is called, the computed point is added to the set of cached values, and the interpolating polynomials are re-estimated each time  \CosmoMC\ tests for convergence.   Figure \ref{fig:useinterp} summarizes the workings of \InterpMC.

\section{Testing  and Results \label{sec:test}}

\begin{table}
\centering
\begin{tabular}[H] {|l |l|}
\hline
\textbf{Shorthand} & \textbf{Variables (+7 Params) }\\ \hline
\hline \hline 
Reference & Concordance Only \\
\hline

Dark energy equation of state   & $w$  \\ \hline
 Curvature & $\Omega_k$ \\
\hline
Tensor:scalar ratio	& $r$ \\
\hline
Running spectral index &$n_{run}$ \\
\hline
Curvature + tensors & $\Omega_k$ + $r$ \\
\hline
Tensors + running & $r$ + $n_{run}$ \\
\hline
Neutrino fraction & $f_{\nu}$ \\
\hline
\end{tabular}
\caption{ The variables used for the trial runs.}
\label{table:variables}

\mbox{}

\begin{tabular}[H] {|l |l|}
\hline
\textbf{Shorthand} & \textbf{Datasets }\\
\hline \hline 
WMAP7 & WMAP7 Likelihood  \\
\hline
gCMB & WMAP7 + Ground based CMB [Define]  \\
\hline
LSS	& WMAP7 + Large Scale Structure \\
\hline
LSS + gCMB & WMAP7 + Large Scale Structure + gCMB  \\
\hline
LSS + gCMB + SN &  WMAP7 + Large Scale Structure + Supernovae \\
\hline
BAO &WMAP7 +  Baryon Acoustic Oscillations    \\
\hline
BAO + SN &  WMAP7 +  Baryon Acoustic Oscillations + Supernovae \\
\hline
\end{tabular}
\caption{ The datasets used for the trial runs.   \label{table:datasets}}
\end{table}

 \begin{figure}[tb]\epsfig{file=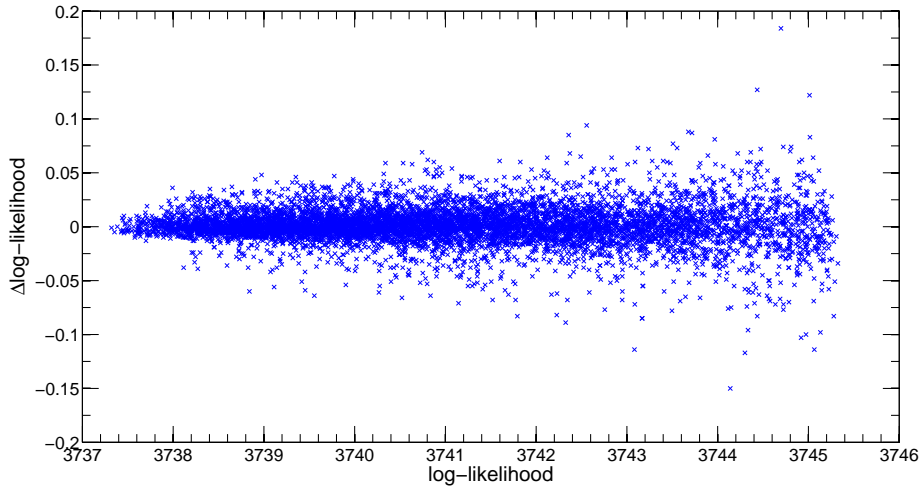,width=14cm} 
        \caption{We plot $\Delta  \log(\cal{L})$, the error in the interpolated log-likelihood, for points in a single representative chain, versus the actual, non-interpolated, $ \log(\cal{L})$.}
	\label{fig:errorsvlog}
	\end{figure}

 \begin{figure}[tb]\epsfig{file=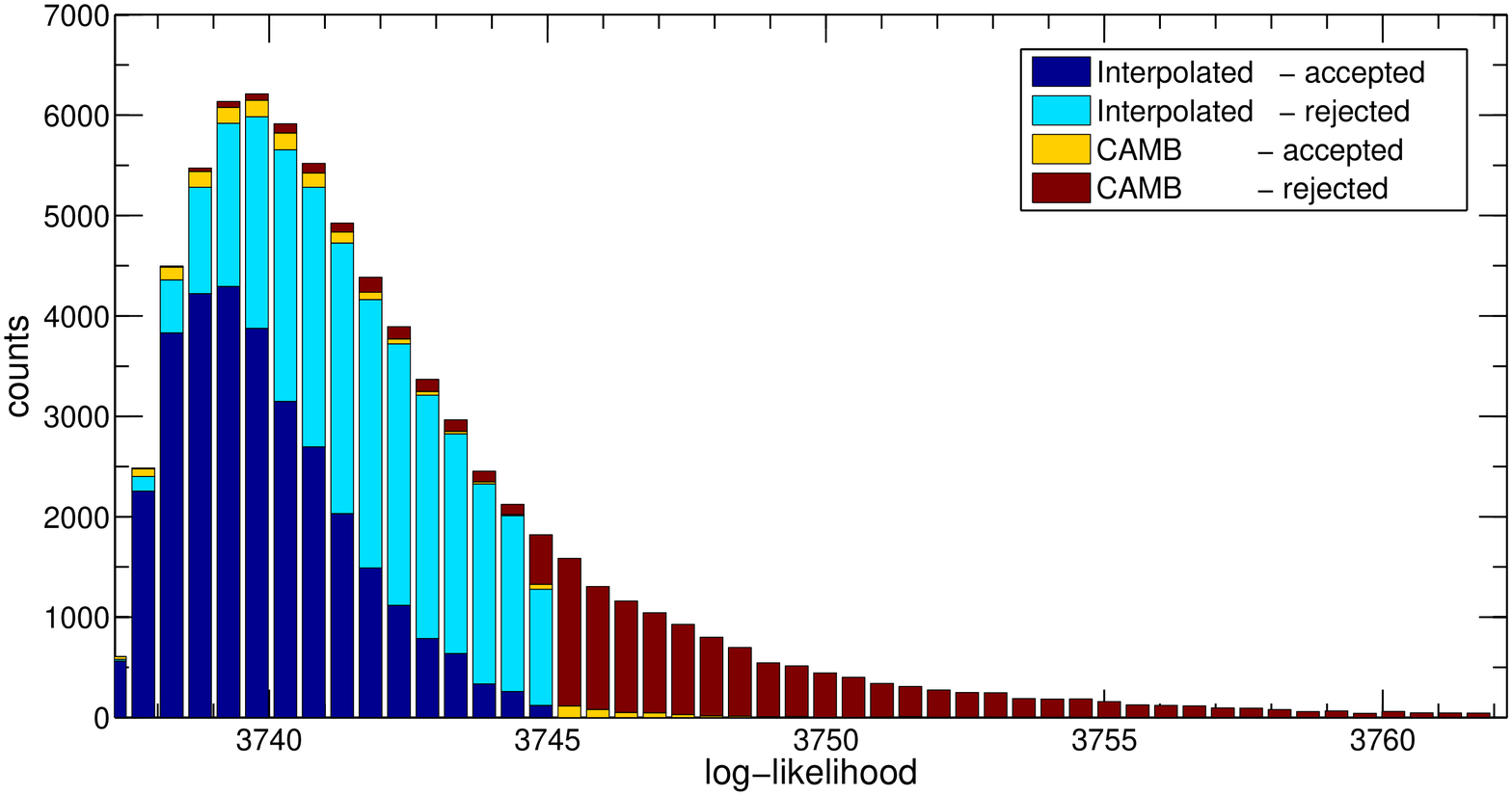,width=6in}
	\caption{ Histogram showing fraction of points in an MCMC chain set for which the likelihood can be obtained via interpolation, once enough data has been accumulated to construct the polynomial.  Points whose interpolated log-likelihood differs by more than 8 (set by the {\tt cut} parameter) are automatically computed from the full likelihood, after running {\sc Camb}.  Points which survive this cut can almost all be obtained directly from the interpolation.	This plot shows results for  $\Lambda$CDM chains, using only the WMAP7 likelihood.}
	\label{fig:histogram}
	\end{figure}

To test \InterpMC\ we ran chains for a number of  parameter sets.  As well as the standard concordance variables, we examined  scenarios with a non-trivial neutrino sector, spatial curvature, tensors, and a running index, as listed in Table~\ref{table:variables}.    For each combination of parameters, we ran chains using only WMAP data, and WMAP in conjunction with other data, including supernovae, large scale structure [LSS] and Baryon Acoustic Oscillation [BAO], as listed in Table~\ref{table:datasets}.   For each choice, we ran both interpolated and conventional chains, and  compared the resulting runtimes and estimated parameter values. We have seven combinations of datasets, and eight combinations of variables, giving a total of 56 distinct estimations.  In each case we found excellent agreement between the interpolated and non-interpolated chains. 

MCMC estimates are inherently stochastic, so there is some variation in the runtimes for otherwise identical estimations, since  the path taken by each chain through the parameter space is a random walk.  However, we found that for a given combination of dataset and model, the ratio between the number of steps and the runtime is roughly constant, while the runtime itself varies by a factor of 2 or more. However, the ratio of runtime/attempted steps is an accurate measure of performance, as illustrated in Figure~\ref{fig:allthree}, and we use this metric to evaluate performance gains associated with \InterpMC.   This constant is a function of the chosen dataset, model, and priors. 
 
We worked with the January 2010 release of \CosmoMC. We used the WMAP7 likelihood code; likelihoods for other datasets are supplied with \CosmoMC. We started with the default settings for the initial proposal matrix with periodic updates enabled, and set {\verb MPI_Converge_Stop } to 0.01. This is more conservative than the default value of 0.03 but yields  smooth two dimensional parameter contours. This choice extends the runtime and thus increases the acceleration induced by the interpolation, since a greater fraction of the chains can use the interpolated likelihood.  The interpolation typically commences well before the changes are close to convergence.

As well as comparing the parameter estimates with and without interpolation, we test the interpolation scheme by calling the regular likelihood and comparing the result to the interpolated value.  The absolute error in the log-likelihood for the interpolated points was on the order of $0.01-0.05$, with increasing error for values further removed from the peak likelihood. In Figure \ref{fig:errorsvlog} we plot the error versus the likelihood.   Given that our parameter estimates overlap well with those returned by the stock version of \CosmoMC, it is empirically clear that this error is not having a dramatic effect on the parameter estimates.   

More quantitively, looking at  Figures \ref{fig:errorsvlog} and \ref{fig:histogram} we see that essentially all the accepted points  in the chains have $\log{\cal L}$ that  differers by less than 10 from the maximum likelihood. Consequently, we do not need a particularly accurate estimate of $\log{\cal L}$ for points far from the peak, since these points are almost always rejected.  Conversely a ``typical'' interpolated point lies has a value of $\log{\cal L}$ that differs by perhaps 3 or 4 from the peak value, and can be interpolated with an error, $ \Delta  \log(\cal{L}) =  \log\cal{L} - \left. \log \cal{L} \right|_{\rm interp.}$ of perhaps $0.02$.  Consequently, the resulting error in $\cal{L}$ is
\begin{equation}
{ \cal{L}} = e^\Delta \exp{[  \left. \log \cal{L} \right|_{\rm interp.}] } \, .
\end{equation}
For the mast majority of interpolated points, $|\Delta| < 0.025$, so the ``exact'' likelihood differs from the interpolated by a few percent.  However, given that MCMC processes are intrinsically stochastic the interpolation error is small enough to be effectively unresolved within the parameter estimates below. 

\begin{table}
\centering
\begin{tabular}[tb] {|l | p{9cm} |l |}
\hline
\textbf{Option} & \textbf{Description}  & \textbf{Default} \\
\hline \hline
\textit{cut} &  Maximal difference between likelihood and maximum likelihood for points included in interpolation.  & 8.0 \\
\hline
\textit{factor} & Factor by which number of free parameters in interpolating polynomial is exceeded by interpolation dataset. & 3.0 \\
\hline
\textit{interp\_order} & Order of interpolating polynomial. & 4 \\
\hline
\textit{fraction\_cut} & Maximal difference between $n$ and $n-1$-th order interpolation. & 0.2 \\
\hline
\textit{do\_interp} &  Set to ``False'' to disable interpolation. & T \\
\hline
\end{tabular}
\caption{Options for running \CosmoMC\ with the interpolation scheme.
\label{table:options}}
\end{table}

 The user-supplied parameters that control the interpolation are added to the   \CosmoMC\ {\tt params.ini} file, and are summarized in Table~\ref{table:options}.    We set the {\em cut\/} parameter to 8, so that interpolation is only used for points where the log-likelihood is within 8 of the maximum value. This encompasses most of the ``peak'' since it excludes points who relative likelihood is $e^8 \approx 3000$ times less than peak value.  Secondly, we require that the second cut difference between the peak and predicted likelihood in the  $n-1^\textrm{th}$ and $n^\textrm{th}$ order interpolations differ by no more than 20\%.   This is not a particularly stringent cut and rejects only a few points, although some of them very close to the peak.  Finally, we used $n=4$, so were comparing $3rd$ and $4th$ order interpolations.     We show the fraction of calls to the likelihood function that are computed via interpolation in Figure~\ref{fig:histogram}.

We ran chains for 56 different combinations of datasets and variables,   using both \InterpMC\ and the original \CosmoMC\ code. We assessed the required runtime by  computing the ratio of runtime / accepted point for each set of chains.    Each estimation was   run using 8 chains on an dual Intel ``Xeon'' cpu node (E5440@2.83 GHz). \InterpMC\ does not add significantly to the MPI overhead,  so \CosmoMC\ continues to be ``embarrassingly parallelizable'', and the interpolation routines take negligible amounts of runtime.

We ran with the parameters defined in the  ``stock'' {\tt params.ini\/} distributed with \CosmoMC, other than changing the convergence parameter to 0.01, in order to ensure that the 2 dimensional parameter constraints were smooth, as noted above.  This allowed the chains to update the proposal matrix as they ran, which shortens the runtime unless the initially supplied covariance matrix is already close to optimal.

\begin{table}
\centering
\scalebox{.85}[.85]{
\begin{tabular}[tb]{|l || l | l | l | l | l | l | l |l|}
\hline
\textit{VARIABLE/DATASET }& 7 Params & $\Omega_k$& $r$ & $n_{run}$ & $\Omega_k$ + $r$ & $r$ + $n_{run}$ & $w$ & $f_{\nu}$ \\ 
\hline \hline
\textbf{WMAP7} & 3.5378 & 2.3794 & 3.5797 & 2.8896 & 2.4169 & 2.4409 & 3.0976 & 3.3019 \\ \hline
\textbf{gCMB} & 3.3525 & 2.4504  & 3.5060 & 3.0079 & 2.7707 & 2.1830 & 2.9112 & 3.8758 \\ \hline
\textbf{LSS}  & 3.3577 & 2.7639  & 2.8640 & 2.0883 & 2.5246 & 2.4469 & 1.8909 & 3.5843 \\ \hline
\textbf{LSS+gCMB} & 3.4446 & 2.9328 & 2.7487 & 2.6698 & 2.7068 & 2.5653 & 2.7886 & 3.6044 \\ \hline
\textbf{LSS+gCMB+SN}& 3.3576 & 2.2650 & 2.3680 & 2.2415 & 2.4163 & 2.8079 & 1.4848 & 3.2155 \\ \hline
\textbf{WMAP+BAO} & 3.5994 & 4.3205  & 3.1864 & 2.4008 & 2.0084 & 2.0426 & 3.0084 & 3.5202\\ \hline
\textbf{WMAP+BAO+SN} & 3.0137 & 3.1614 & 3.1986 & 2.2648 & 2.0851 & 2.3197 & 3.4321 & 3.3011\\ \hline
\end{tabular}}
\caption{Performance improvement for \InterpMC, relative to standard \CosmoMC\ for different choices of parameters and dataset.  }
\label{table:ratios}
\end{table}

\begin{figure}[tb]\epsfig{file=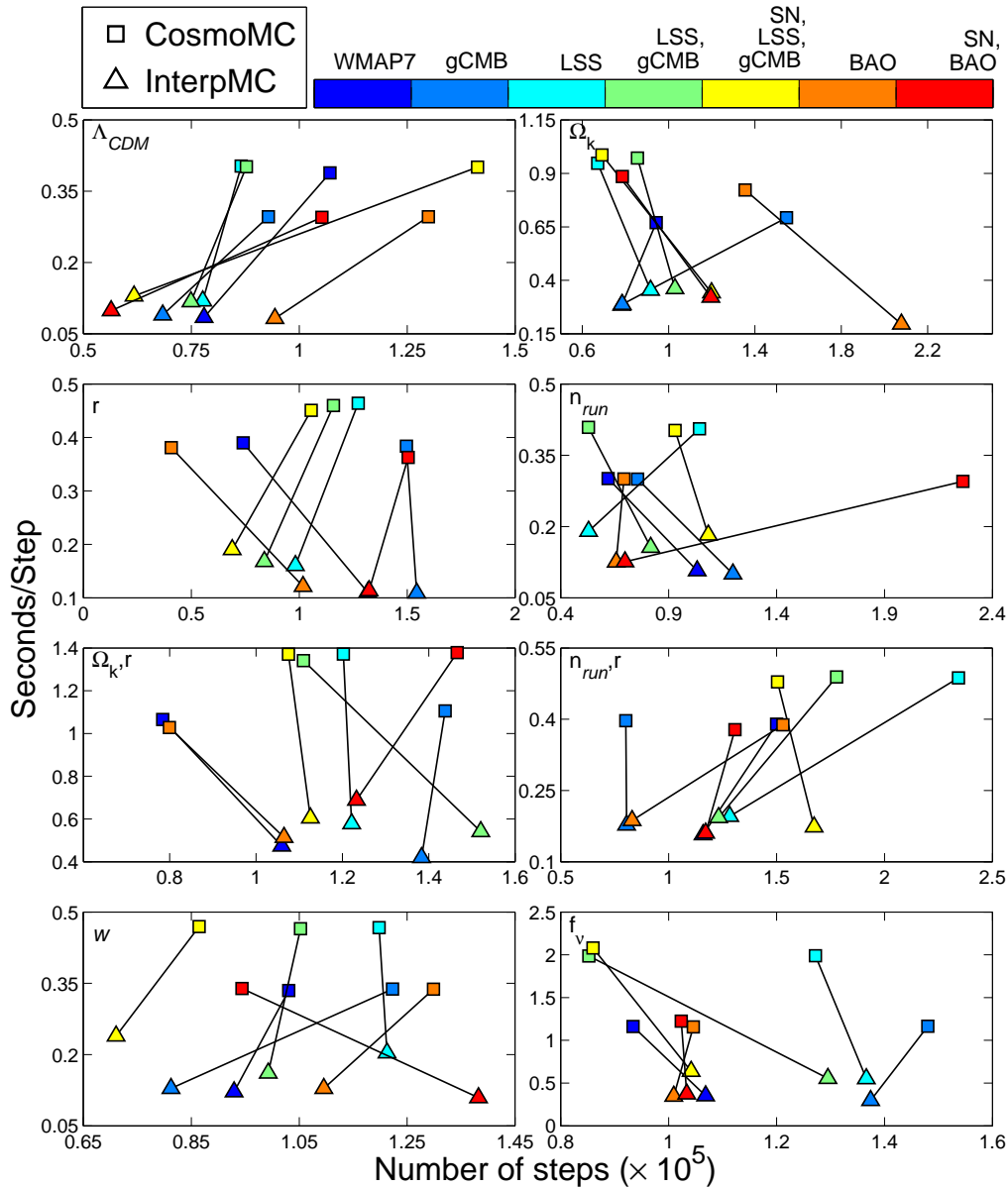,width=6in}
\caption{Runtime per (attempted) step for \InterpMC\ (triange) and stock \CosmoMC\  (square) runs versus the total number of steps,  for each parameter set. Black lines connect runs preformed with  \InterpMC\  and \CosmoMC. The required number of steps varies stochastically even for otherwise identical runs, and there is no systematic difference between the number of steps required by \InterpMC, relative to the stock code.  }
\label{fig:runtimesvariables}
\end{figure}
   
 \begin{figure}[thb]
 \epsfig{file=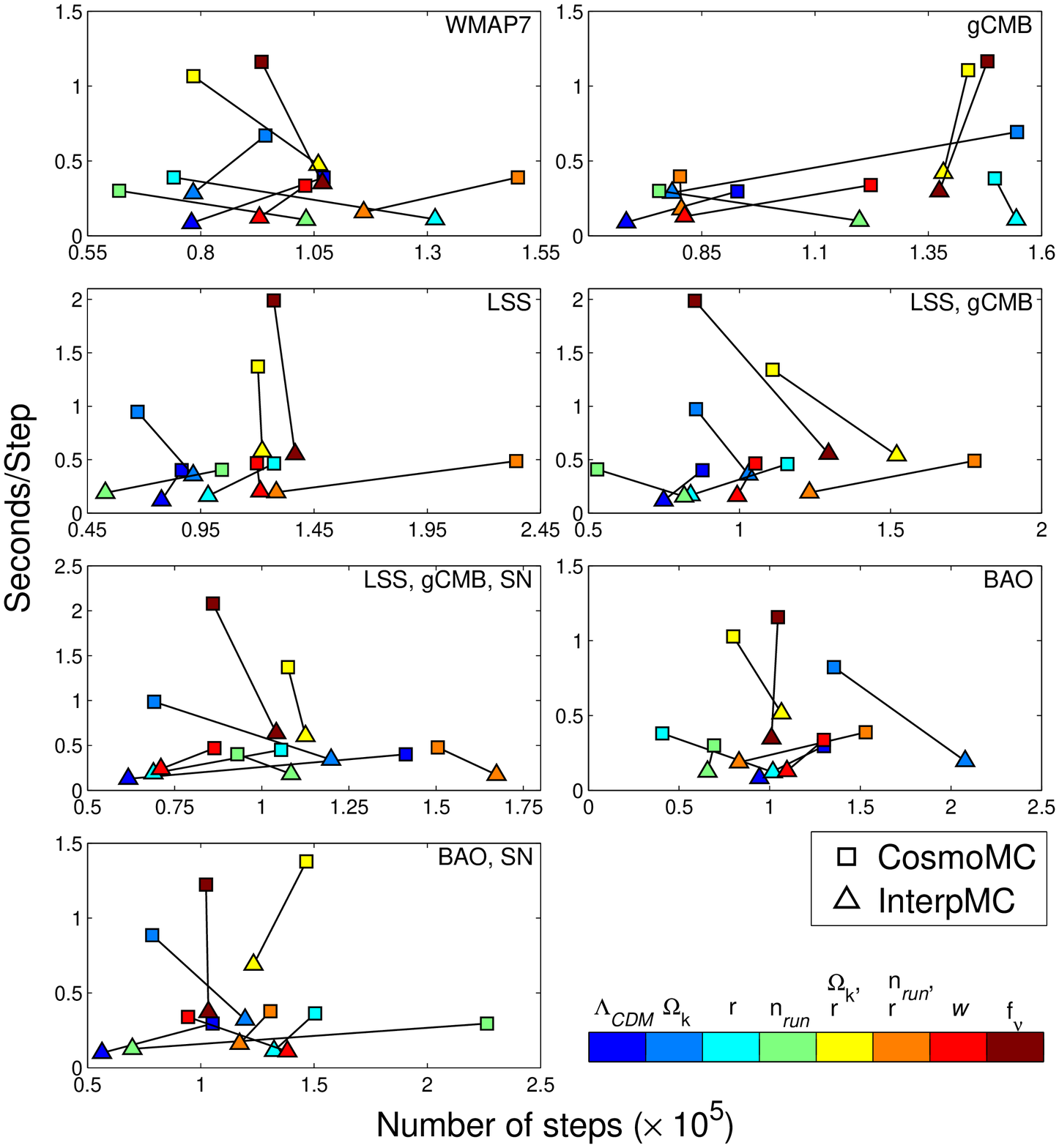,width=6in}
\caption{Runtime per (attempted) step for \InterpMC\ (triange) and stock \CosmoMC\  (square) runs versus the total number of steps,  for each combination of datasets.  Black lines connect runs preformed with  \InterpMC\  and \CosmoMC.    }
\label{fig:runtimesdatasets}
   \end{figure}

\begin{figure}[tb]\epsfig{file=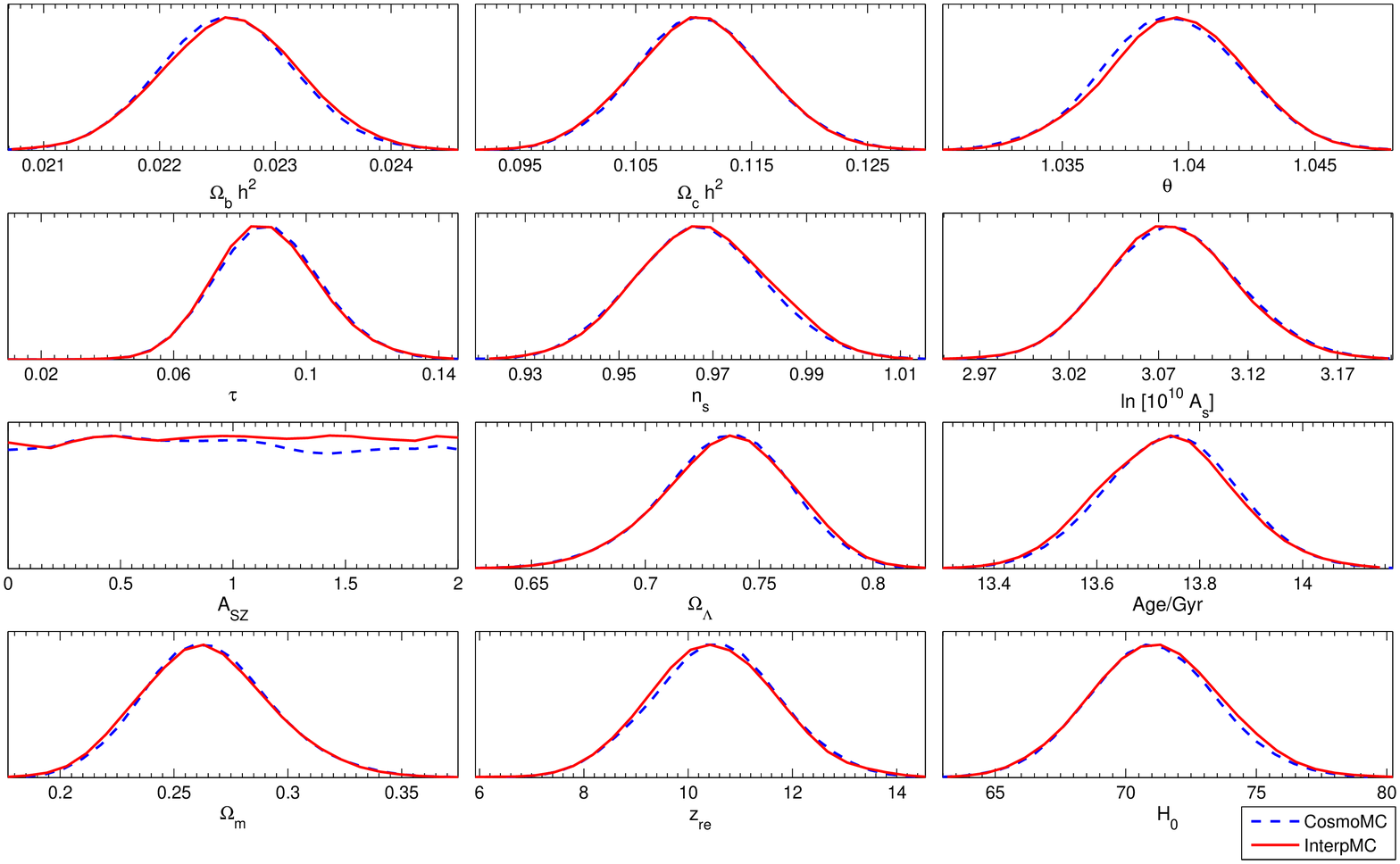,width=5.8in}
\epsfig{file=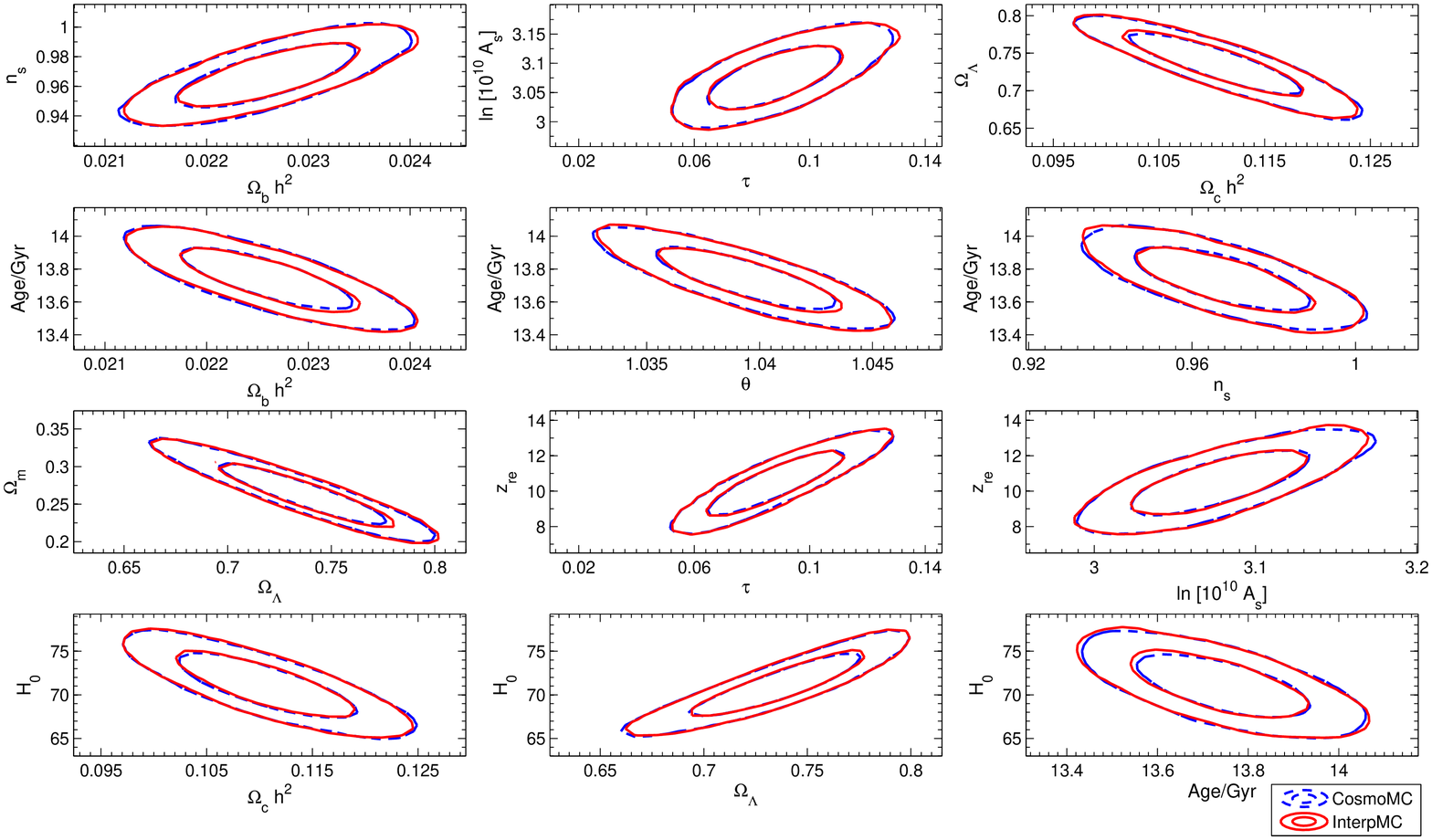,width=5.8in}  
	\caption{We show 1D and 2D marginalized likelihood plots for the reference 7 parameter run using only WMAP, comparing results from an unmodified ComsoMC run, and results from  \InterpMC.}
	\label{fig:7params}
	\end{figure}

\begin{figure}[tb]\epsfig{file=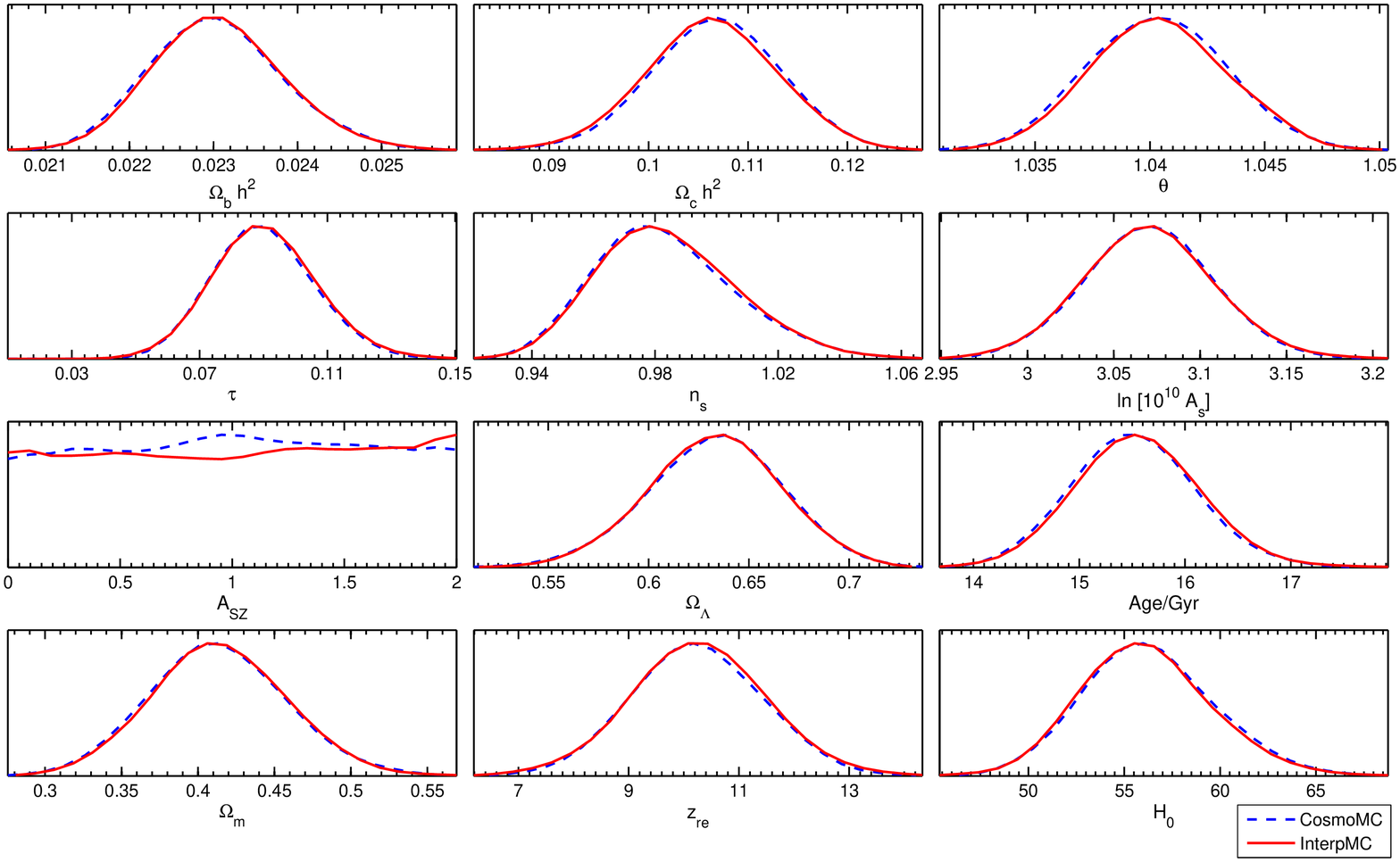,width=5.8in}
\epsfig{file=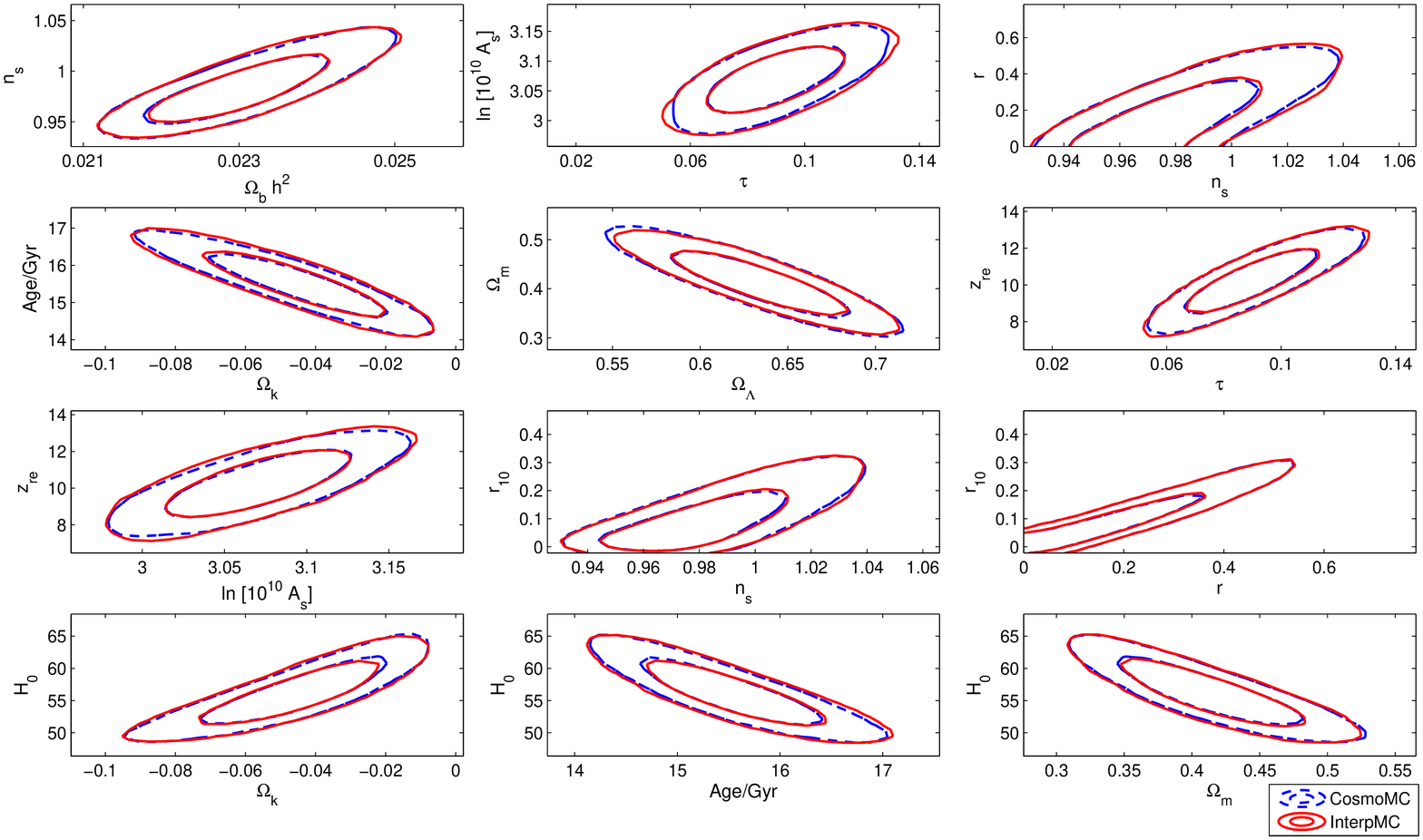,width=5.8in}
	  \caption{We show 1D and 2D marginalized likelihood plots for the spatial curvature and tensors ($\Omega_k$ and $r$) using BAO and SN, comparing results from an unmodified ComsoMC run,  and results from    \InterpMC.  }
	  \label{fig:run74}
	  \end{figure}

The performance of \InterpMC\ is shown in Table~\ref{table:ratios}, and we see that the typical improvement in runtime is between 2  and 4.  We look at models with up to 9 free parameters, and do not see  an obvious correlation with between the performance gain and the total number of parameters.    Figure~\ref{fig:runtimesvariables}  shows the performance of \InterpMC\ as a function of the chosen set of free parameters, while  while Figure~ \ref{fig:runtimesdatasets}  shows the performance as a function of the datasets being employed for the estimation.   Finally, looking at representative parameter estimates shown in Figures~\ref{fig:7params} and \ref{fig:run74} we see that the likelihood contours produced by \CosmoMC\ and \InterpMC\ are effectively identical; the only difference between them is in the runtime required to produce them.

%

\section{Discussion \label{sec:disc}}

We describe an approach to  reducing the runtime required for cosmological parameter estimation, based on interpolating cached likelihood values computed with a single set of  MCMC chains. An implementation of this scheme, \InterpMC, is available as a patch to the standard MCMC package \CosmoMC.   This analysis focuses on  cosmological parameter estimation, but the fundamental approach is applicable to other problems where the likelihood is a smooth function of the free parameters, and computationally intensive to evaluate.

We should ask {\em why\/} this approach works.    Firstly, and unlike methods based on a precomputed training set (unless one has a very good idea of where to look, which is really only possible after an estimation has been performed), Markov Chains primarily evaluate points in regions of high likelihood, so our dataset is naturally weighted towards those points that will be frequently sampled during the estimation. Secondly, while MCMC estimates are much more efficient that a brute force search of the parameter space, they  have no ``memory'' -- their behavior is determined solely by the current location of the chain in parameter space.   Any interpolation scheme, whether generated on the fly  or via a precomputed training set is, to some extent, exploiting the global properties of the likelihood function. In particular, cosmological likelihoods are typically smooth functions of the free parameters we are seeking to estimate. This greatly facilitates the construction of an interpolating polynomial, but this information is not exploited by the MCMC algorithm itself.  As a side note, we would point that cosmological parameter estimation already makes substantial use of  interpolation:  CAMB directly computes only a subset of the underlying Fourier modes and  the $C_\ell$ values and then fits the intervening points. Turning off  this interpolation significantly increases the runtime of CAMB. Consequently,  \InterpMC\ is extending the use of interpolation in cosmological parameter estimation, rather than introducing it for the first time.

As presently implemented, the efficiency  gain yielded by \InterpMC\ is typically between  2 and  4.   Our primary goal here is  simply to demonstrate the feasibility of this approach. We have deliberately  employed conservative choices of the adjustable parameters in order to demonstrate that this approach works without careful tuning.  Our experiments have shown that  choosing more   aggressive settings for the free parameters in \InterpMC\ can noticeably increases the efficiency gain delivered by \InterpMC\ without any modification to the underlying algorithm.    Further, the interpolation requires a very small fraction of the total runtime, so even in the worst case scenario it cannot significantly slow \CosmoMC, relative to a standard run. 

We can see several specific algorithmic changes that may significantly improve the performance of \InterpMC.  Firstly, we currently fit to all combinations of all free parameters, up to order $n$. However, many of these parameters are largely uncorrelated, and a more intelligent (but still automated) fitting procedure would focus on the subset of parameter combinations  which make a nontrivial contribution to the fit, allowing the chains to begin making use of interpolated likelihoods at an earlier point in the run.   Separately, the current interpolation is expressed in terms of the (normalized) free parameters in the chains.  However, we can always choose a basis in which the free parameters are uncorrelated at second order, so that the coefficient of $x_i x_j$ in the interpolating polynomial is proportional to $\delta_{ij}$. In this case, the higher order coefficients (e.g.  $c_{ijk} x_i x_j x_k$) would reflect the couplings between variables whose covariance vanishes at second order, and will likely have a smaller number of nontrivial terms than the corresponding polynomial written in terms of the unrotated variables.  Further,  we could move to a scenario in which some combinations of variables are interpolated at higher order than others.   Finally, while we have focused on the ability to avoid precomputation, it would be simple to save the computed likelihoods to ``seed''  a future run\footnote{This might be useful in situations where the likelihood computation was unchanged, but the allowed parameter ranges were altered, or the chains were to be extended for better convergence.}, or to accelerate the subsequent computation of Bayesian evidence.

Our goal here has been to interpolate the likelihood in a way that is essentially invisible to the user and does not modify the underlying MCMC algorithm in any way.  However, we have effectively demonstrated that we can construct an accurate interpolation to the likelihood for a wide range of cosmological datasets and models, while expending less computational effort than that which is required for a full parameter estimation.  The interpolating polynomial effectively yields the functional form of the likelihood in a substantial volume surrounding the peak. Consequently,  the existence of a closed-form algebraic expression for the likelihood may allow us to pursue parameter estimation techniques that make direct use of this information, beyond standard MCMC techniques.

 It is clear that the computational cost of cosmological parameter estimation will continue to rise, particularly once data from Planck becomes available. Not only will this dataset include information at smaller angular scales than WMAP, the signal to noise will also be substantially improved,  requiring a more accurate evaluation of the theoretical $C_\ell$, via CAMB. The computational cost of CAMB rises rapidly with the precision of the $C_\ell$, whereas the interpolation algorithm runs in negligible amounts of time.  Further, both the likelihood code and CAMB itself undergo frequent modifications and updates, each of which would require a precomputed training set to be regenerated from scratch. In particular, the likelihood may undergo many changes as a new dataset is analyzed and reduced. Consequently \InterpMC's avoidance of trainings sets that are generated separately from the parameter estimation  process guarantees that it will reduce the overall computational cost of parameter estimation, rather than just shifting it into the computation of the training set.
 
To summarize, we have presented an initial implementation of an interpolation driven approach to accelerating MCMC  cosmological parameter estimations, and shown that it can produce accurate results while reducing the resulting computational expenditure by at least a factor of two with a simple ``proof of concept'' implementation of this algorithm.  Further, we have identified specific improvements to this approach that promise to substantially improve our current performance gains, and could be implemented in a production-ready version of \InterpMC.

 \acknowledgments 
 RE is supported in part by the United States Department of Energy, grant DE-FG02-92ER-40704 and by an NSF Career Award PHY-0747868. This work was supported in part by the facilities and staff of the Yale University Faculty of Arts and Sciences High Performance Computing Center.

\bibliographystyle{h-physrev3}
\bibliography{interp}

\end{document}